\documentclass[12pt]{article}
\usepackage{theorem}
\theorembodyfont{\upshape}

\newtheorem{theorem}{Theorem}[section]
\newtheorem{lemma}[theorem]{Lemma}

\newtheorem{definition}[theorem]{Definition}

\newcounter{abc}[equation]

\newcommand*{\picdirrectory}{}
\newcommand*{\fig}[3]{
        \begin{figure}[!h!t]
	\begin{center}
        \includegraphics{\picdirrectory#1}
	\end{center}
        \caption{\footnotesize #2}
        \label{#3}
        \end{figure}}

\usepackage{amssymb,amsfonts}
\usepackage{graphics}

\makeatletter
\@addtoreset{equation}{section}
\makeatother

\def\ZZ{{\mathbb Z}}
\def\NN{{\mathbb N}}
\def\RR{{\mathbb R}}

\def\indic{{\bf 1}}

\newenvironment{eqs}{\begin{equation}\left\{\begin{array}{ll}}{\end{array}\right.\end{equation}}
\newenvironment{eqe}{\begin{equation}\begin{array}{rl}}{\end{array}\end{equation}}
\newenvironment{eqc}{\begin{equation}\begin{array}{c}}{\end{array}\end{equation}}
\newcommand*{\be}{\begin{eqs}}
\newcommand*{\ee}{\end{eqs}}
\newcommand*{\bee}{\begin{eqe}}
\newcommand*{\eee}{\end{eqe}}
\newcommand*{\boe}{\begin{eqc}}
\newcommand*{\eoe}{\end{eqc}}
\newenvironment{proof}[1]{{\noindent{\sc Proof#1:}}}{\begin{flushright}$\square$\end{flushright}}

\newcommand*{\sumu}{\displaystyle\sum}

\renewcommand*{\epsilon}{\varepsilon}

\newcommand{\xit}{{\xi}}

\newcommand{\taut}{{\tau}}

\newcommand{\llt}{{l}}
\newcommand{\rrt}{{r}}
\newcommand{\GRt}{{GR}}
\newcommand{\Pit}{{\Pi}}

\newcommand{\notto}{{\to\kern-1.15em/\;\;}}
\newcommand{\nottot}{{\tot\kern-1em/\;\;}}
\newcommand{\tot}{{\rightsquigarrow}}

\renewcommand{\picdirrectory}{bsd.pic/}

\begin{document}
\pagestyle{myheadings} \markright{A discrete Bak-Sneppen model} 
%\markleft{Ronald Meester and Dmitri Znamenski}
\setlength{\textheight}{21cm}
\title{{\bf Non-triviality of a discrete Bak-Sneppen evolution model}}
\author{Ronald Meester\footnote{ Faculty of Exact Sciences, Free University Amsterdam, de Boelelaan 1081,
1081 HV Amsterdam, The Netherlands; telephone +31-20-444-76-72, fax +31-444-76-53, e-mail: 
dznamen@cs.vu.nl} 
and Dmitri Znamenski$^*$}
\maketitle
\begin{abstract}
Consider the following evolution model, proposed in \cite{BS} by Bak and Sneppen. Put $N$ 
vertices on a circle, 
spaced evenly. Each vertex represents a certain species. We associate with each vertex a 
random 
variable,
representing the `state' or `fitness' of the species, with values in $[0,1]$. The dynamics 
proceeds as 
follows. Every discrete time step, we choose the vertex with minimal fitness, and assign 
to this vertex, and to its two neighbours, three new independent fitnesses with a uniform 
distribution on $[0,1]$. 
A conjecture of physicists, based on simulations, 
is that in the stationary regime, the one-dimensional marginal distributions of the 
fitnesses converges, 
when $N \to \infty$, to a uniform distribution on $(f,1)$, for some threshold 
$f<1$.

In this paper we consider a discrete version of this model, proposed in \cite{BK}.
In this discrete version, the fitness of a vertex can be either $0$ or $1$. The system 
evolves according 
to the following rules. Each discrete time step, we choose an arbitrary vertex with 
fitness 0. If all 
the vertices have 
fitness 1, then we choose an arbitrary vertex with fitness 1. Then we update 
the fitnesses of this vertex and of its two neighbours by three new 
independent fitnesses, taking value 0 with probability $0<q<1$, and 1 with probability 
$p=1-q$. We 
show that if $q$ is close enough to one, then the mean average fitness in the stationary 
regime 
is bounded away from $1$, uniformly in the number of vertices. This is a small step in the 
direction of 
the conjecture mentioned above, and also settles a conjecture mentioned in \cite{BK}.

Our proof is based on a reduction to a continuous time particle system.


\end{abstract}

\section{Introduction}

The Bak Sneppen model, introduced in \cite{BS}, has received a lot of attention in the 
literature, see for instance 
\cite{B}, \cite{J}, \cite{BSm}, \cite{BSl}, \cite{BSh}, \cite{BSx}, \cite{BSv} and \cite{BSs}. In 
\cite{B}, it is described how Bak and Sneppen were looking for a 
simple  mathematical model which was supposed to exhibit evolutionary behaviour, and which was 
also supposed to fall 
into the class of processes showing self-organised critical behaviour. For physicists, 
self-organised critical behaviour refers to power law decay of temporal and spatial 
quantities, without fine-tuning of parameters.
After many attempts, Bak and Sneppen arrived at the following process.

Think of a system with $N$ species. These species are represented by $N$ vertices on a 
circle, evenly 
spaced. Now each of these species is assigned a so called `fitness', and in this model, 
the fitness 
is a number between 0 and 1. The higher the fitness, the better chance of surviving the 
species has. 
The dynamics of evolution is modelled as follows. Every discrete time step, we choose the 
vertex with 
minimal 
fitness, and we think of the corresponding species as disappearing completely. This 
species is then
replaced by a new one, with a fresh and independent fitness, uniformly distributed on 
$[0,1]$. So far, 
the dynamics does not have any 
interaction between the species, and does not result in an interesting process. 
Interaction is 
introduced by also replacing the two neighbours of the vertex with lowest fitness by new 
species 
with independent fitnesses. This interaction represents co-evolution of related species: if a certain 
species becomes extinct, this has an effect on other species as well. The neighbour interaction makes the 
model very interesting 
from a mathematical point of view.

It is extremely simple to run this model on a computer. Simulations then suggest the 
following 
behaviour, for large $N$ (see \cite{J} and \cite{B} for simulation results). It appears 
that the 
one-dimensional marginals are uniform (in the limit for $N \to \infty$) on $(f,1)$ for 
some $f$ whose 
numerical value is supposed to be close to 2/3. This threshold value $f$ is the basis for 
self-organised 
critical behaviour, according to \cite{BS}, \cite{B} and \cite{J}. Since in the limit 
there is no mass 
below $f$, one can look at so called avalanches of fitnesses below this threshold: start 
counting at the 
moment that there is one fitness below $f$ and wait until all fitnesses are above $f$ 
again. The random 
number of updates, for instance, counted this way, is suppose to follow a power law, 
and there is no fine-tuning of parameters.

It is a challenge to prove any of the above statements. Note that in order to prove power law behaviour,
one should first prove the existence of the threshold $f$ with the property that in the limit for
$N \to \infty$, all one-dimensional marginals are concentrated on $(f,1)$. Indeed, one can define 
avalanches corresponding to other thresholds as well, but it is not expected that these avalanches have 
power law behaviour. This is only expected (and observed) for the self-organised threshold $f$. 
Therefore, this should be the starting point of a rigorous mathematical analysis of the model. 

Simple as the model may appear, it turns out to be very difficult to say anything at all about the 
limiting one-dimensional distributions. It is therefore natural to try to prove a similar result in a 
simpler model. In this light, we have chosen to study a discrete version of the model, which was proposed 
in
\cite{BK}, and which can be described as follows. Fitnesses of species can now only be 0 or 1. The 
dynamics in this simpler model 
proceeds as  follows: at every discrete time step, we choose an arbitrary vertex with fitness 0. If 
there is no such  vertex, then we choose an arbitrary vertex with fitness 1. We update 
the fitnesses of this vertex, and of its two neighbours, by three new 
independent fitnesses, taking value 0 with probability $0<q<1$, and 1 with probability 
$p=1-q$. This 
process is called the BS process in this paper. We 
show that if $q$ is close enough to one, then the mean average fitness in the stationary 
regime  is bounded away from $1$, uniformly in the number of vertices. This is a small step in the 
direction of the conjecture mentioned above, and answers a question which was posed in \cite{BK}.

It should be noted that this discrete version of the model does not show self-organised critical 
behaviour. Nevertheless, we think that understanding of the discrete model also increases our
understanding of the original model, if not in a technical sense, certainly in a conceptual sense. 
Admittedly, the discrete version suggested here and in \cite{BK} is only one out of many possible 
discrete versions, but we see no reason to complicate matters unnecessarily 
by choosing more complicated discrete versions. The reader will notice that the proof of our main result 
is already quite complicated. 

In order to state our main result, here follows some notation. As before, the number of 
vertices is 
denoted by $N$, and we denote by $\eta_N(n)_i$ the state of the $i$-th vertex after $n$ 
updates of the process. We will prove the following result.

\begin{theorem}
\label{P1}
If $q$ is close enough to one, then there exists $c_q>0$ such that for any $N\in\NN$ and 
$i \in 
\{1,\ldots, N\}$,
\boe
\label{eq25}
\lim\limits_{n\to\infty}
P(\eta_N(n)_i=0)\ge c_q.
\eoe
\end{theorem}
Note that $\lim\limits_{n\to\infty}P(\eta_N(n)_i=0)$
exists, because $\eta_N(n)$ is a finite 
state, irreducible and aperiodic Markov chain.

In the next section, we reduce the problem to a problem in a continuous time, monotone 
particle system. 
In this system we will be able to prove results uniformly in $N$, by exhibiting graphical 
representations and an infinite space version of the particle system.

\section{Reduction to a monotone continuous time process.}

In this section we define a useful
continuous time stochastic process $\xit(t)$, independent of $N$.
We construct the process $\xit(t)$ via
a graphical representation.
The graphical representation $\GRt$ is a random graph on the
space-time diagram $\ZZ\times\RR^+$.
We define $\GRt$ via
a set of independent so called {\em bundles} $\{\Pit_k\}_{k\in\ZZ}$,
where each bundle $\Pit_k$ consists of eight independent
Poisson processes on $\RR$,
$$
\Pit_k=\{\Pit^{000}_k,\Pit^{001}_k,\dots,\Pit^{110}_k,\Pit^{111}_k\},$$
with parameters $q^3/(1-q^3),q^2p/(1-q^3),\dots,p^2q/(1-q^3),p^3/(1-q^3)$
respectively. (We use the factor $1/(1-q^3)$ to rescale time
in a convenient way, as will become clear later.)
For each process $\Pit^{\sigma_1,\sigma_2,\sigma_3}_k$ we
perform the following procedure.
At $i$-th arrival 
$\taut^{\sigma_1,\sigma_2,\sigma_3}_{k,i}$ of 
$\Pit^{\sigma_1,\sigma_2,\sigma_3}_k$, $i\in\ZZ$,
we draw arrows in $\ZZ\times\RR$ from
$(k,\taut^{\sigma_1,\sigma_2,\sigma_3}_{k,i})$ to
$(k-1,\taut^{\sigma_1,\sigma_2,\sigma_3}_{k,i})$, iff $\sigma_1=0$,
and from
$(k,\taut^{\sigma_1,\sigma_2,\sigma_3}_{k,i})$ to
$(k+1,\taut^{\sigma_1,\sigma_2,\sigma_3}_{k,i})$, iff $\sigma_3=0$.
We draw a $*$ in $\ZZ\times\RR$ at every
$(k+j,\taut^{\sigma_1,\sigma_2,\sigma_3}_{k,i})$ 
with $\sigma_j=1$, $j=-1,0,1$.
We say that $(x,t_1)$ is {\em connected }to $(x,t_2)$ by a {\em time segment},
if $t_1<t_2$ and there are no $\ast$'s on $(x,t)$, $t\in[t_1,t_2)$.
An {\em open path} $\gamma(t)|_{t_1}^{t_2}$ 
with {\em trace} $(x_0,s_0),\dots,(x_n,s_n)$
is a map from $[t_1,t_2]$ to $\ZZ$ such that
$t_1=s_0<\cdots<s_n=t_2$,
$\gamma(t)=x_i$, for $t\in[s_i,s_{i+1})$, $0\le i<n$,
$\gamma(t_2)=x_n$, and every pair
$(x_i,s_i)$, $(x_{i+1},s_{i+1})$ is connected either by
a time segment or by an arrow. 
For any finite $B\subset\ZZ$, $x,y\in B$, and $t_1\le t_2\in\RR$,
write $(x,t_1)\tot(y,t_2)$ in $\GRt|_B$,
if there exists an open path $\gamma(t)|_{t_1}^{t_2}$ in $\GRt$ 
with $\gamma(t_1)=x$, $\gamma(t_2)=y$, and $\gamma(t)\in B$, 
for all $t\in[t_1,t_2]$.

\fig{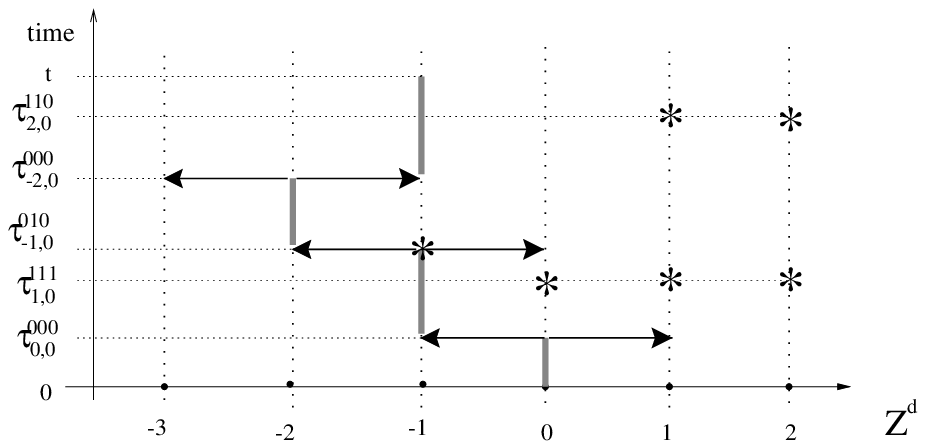}{The graphical representation $\GRt$.
$(0,0)$ is connected to $(-1,t)$ by an open path.
The trace is $(0,0)$, $(0,\taut^{000}_{0,0})$,
$(-1,\taut^{000}_{0,0})$,
$(-1,\taut^{010}_{-1,0})$, $(-2,\taut^{010}_{-1,0})$,
$(-2,\taut^{000}_{-2,0})$, $(-1,\taut^{000}_{-2,0})$
$(-1,t)$.}{f4}

For any finite $A\subseteq B\subset\ZZ$ and $t,s\ge0$
we denote by $\xit^{(A,t)}_B(s)$ the
random subset of vertices $x\in B$ such that there exists
$y=y(x)\in A$, and $(y,t)\tot(x,t+s)$ in $\GRt|_B$. 

The definition of $\xit^{(A,t)}_B(s)$ via the existence
of certain open paths implies a number of useful properties.
The first is monotonicity:
\boe
\label{eq40}
\xit^{(A,t)}_B(s)\subseteq
\xit^{(C,t)}_D(s),\,s\ge 0
\qquad\mbox{if }A\subseteq C,\,B\subseteq D,
\eoe
The second is the semigroup property:
\boe
\label{eq55}
\begin{array}{ll}
\mbox{(a) }&
\xit^{(\{\emptyset\},0)}_B(t)=\{\emptyset\},\qquad t\ge0,\\
\mbox{(b) }&
\xit^{(\xit^{(A,t)}_B(s_1),t+s_1)}_B(s_2-s_1)=
\xit^{(A,t)}_B(s_2),
\qquad\mbox{if }A\subseteq B, 0\le s_1\le s_2.
\end{array}
\eoe

The monotonicity property~(\ref{eq40}) allows us to
define the process $\xit^{(A,t)}_B(s)$ for any $A\subseteq B\subseteq\ZZ$:
\boe
\label{eq42}
\xit^{(A,t)}_B(s)
=\lim\limits_{
\begin{array}{c}
\scriptstyle{B'\uparrow B},\\
\scriptstyle{B' \mbox{ finite}}
\end{array}
}
\xit^{(A\cap B',t)}_{B'}(s).
\eoe
Note that due to the monotonicity property~(\ref{eq40}),
the limit at the r.h.s.\ is independent of the sequence
$B'\uparrow\ZZ$.
The process $\xit(t)$ is now defined as follows:
\boe
\label{eq69}
\xit(t)= \xit^{(\ZZ,0)}_{\ZZ}(t),\qquad t\ge 0.
\eoe

We now extract the BS process $\eta_N(n)$ from the 
graphical representation $\GRt$ as to have $\eta_N(n)$
and $\xit(t)$ defined on the same probability space.

The $N$ vertices  are labeled by $\Lambda(N)=\{-N'-1,\ldots, N''+1\}$, 
where $N'+3+N''=N$, $N''-1\le N'\le N''$, and the observation 
site is labeled by 0. We define $l(i)$ and 
$r(i)$ to be the left and right neighbours of $i$, respectively, with appropriate boundary 
conditions:
$$
l(i)=\left\{
\begin{array}{rl}
i-1,&\; i\in[-N',N''+1],\\
N''+1,&\; i=-N'-1,
\end{array}\right.
$$
$$
r(i)=\left\{
\begin{array}{rl}
i+1,&\; i\in[-N'-1,N''],\\
-N'-1,&\; i=N''+1.
\end{array}\right.
$$

A state of the BS process is determined by the subset
of the vertices in state 0. Thus the state space ${\cal S}_N$ of the
BS process consists of the all subsets of $\Lambda(N)$. If we denote 
the state of a site $i$ in a configuration $\eta\in {\cal S}_N$ 
by $\eta_{i}$, then we have an identity
\boe
\label{eq26}
\eta_{i}=\indic\bigg\{i\notin \eta\bigg\}.
\eoe
This identity is natural because the 0's play
the `active' role in the dynamics of BS process. It is possibly also slighty
inconvenient for mathematicians, who are used to work with subsets
of sites in state 1, (like in the contact process or in the 1-dim 
sendpile model, ect.). Indeed, for those processes we would have 
$\eta_{i}=\indic\bigg\{i\in \eta\bigg\}$.
Nevertherless, we will not reverse the 
roles of 1's and 0's, and will work with~(\ref{eq26}), 
because of the conventional definition of the BS-model.

We will now extract from $\GRt$ two independent sequences
of random variables $U=(U_j)$, $V=(V_j)$, and then we
will define the BS process in terms of those sequences.
Let $\Pit(N)$ denote the superposition of all the Poisson processes 
associated to the vertices in $\Lambda(N)$, i.e., with abuse of notation,
$$
\Pit(N)=\bigcup\limits_{k\in\Lambda(N)}\bigg(
\Pit^{000}_k\cup\Pit^{001}_k\cup\cdots\cup\Pit^{110}_k\cup\Pit^{111}_k
\bigg).
$$
Then $\Pit(N)$ is a Poisson process on $\RR$ with intensity 
$N/(1-q^3)$.
Denote by 
\boe
\label{eq70}
\taut(N)=(\taut_1,\taut_2,\dots)
\eoe
the arrivals of $\Pit(N)$ after time zero. For every 
$j\in\NN$ there exists, with probability one,
a unique $U_j\in\Lambda(N)$ and $V_j\in\{0,1\}^3$ 
such that $\taut_j$ is the arrival of 
$\Pit^{V_j}_{U_j}$. 
It is clear that $U$, $V$ and $\taut(N)$
are independent and each consists 
of i.i.d. random variables. 
Note that $U_j$, $j\in\NN$ is uniformly distributed, that is, 
$P(U_j=i)=1/N,\, i\in\Lambda(N).$
The sequence $U$ will be used as a sequence of random 
vertices-canditates for the update procedure. 
The distribution of $V_j$ is simply the joint distribution of
three independent Bernoulli random variables, 
taking value 0 with probability $q$ and 1 with probability $1-q$.
The sequence $V$ will be used to determine the states
of the vertices after the updates.

We will define $\eta_N(n)$ inductively via the (random) increasing sequence 
$(j_n)\subset \NN$.
Let $j_0=0$, 
$\eta_N(0)=\emptyset$. 
Let $n\in\NN$, and suppose that $j_{n-1}$ and  $\eta_N(n-1)$ are already defined.
If $\eta_N(n-1)=\emptyset$, then $j_n= j_{n-1}+N+1$,
$\eta_N(n)=\{U_{j_n}\}$, i.e. we skip
$N$ elements
of the sequences $U$ and $V$, and then restart our process
from the site $U_{j_n}$. The reason to skip $N$ elements is that we want to
have the following property: the more particles
in state 1 we have in $\eta_N(n-1)$, the more elements
of $U$ and $V$, in mean, we skip to define $\eta_N(n)$.
We will use this property later, in the proof of
Lemma~\ref{L8}. If $\eta_N(n-1) \neq \emptyset$, we wait until we choose a vertex in state 
0:
$$
j_n=\min\bigg\{j>j_{n-1}\,
|\,U_j\in\eta_N(n-1)\bigg\},
$$
and change the state of site $U_{j_n}$ and its
neighbours $l(U_{j_n})$ and $r(U_{j_n})$ according
the value of $V_{j_n}=(\sigma_1,\sigma_2,\sigma_3)$, i.e.
$$
\eta_N(n)_i=\left\{
\begin{array}{ll}
\sigma_1,&i=l\bigg(U_{j_n}\bigg),\\
\sigma_2,&i=U_{j_n},\\
\sigma_3,&i=r\bigg(U_{j_n}\bigg),\\
\eta_{N}(n-1)_i,&\mbox{otherwise}.
\end{array}\right.$$
This finishes the construction of the process $\eta_N(n)$.

\medskip
We will now introduce an `intermediate' continuous time process $\xit^R_N(t)$:
$$
\xit^R_N(t)=\left\{
\begin{array}{ll}
\eta_N(0),&\qquad t\in[0,\taut_1),\\
\eta_N(n(j)),&\qquad t\in[\taut_j,\taut_{j+1}), j\ge1,
\end{array}\right.$$
where $(n(j))$ is defined as
$$
n(j)=\max\bigg\{n\in\NN\,
:\,j_n\le j\bigg\}.
$$
It is clear that $\xit^R_N(t)$ is a continuous time
Markov chain on ${\cal S}_N$. Observe that the processes $\xit^R_N$ and $\eta_N$ are 
related via
a random time change. If there are many 1's around, then we typically skip more steps, so 
1's tend 
to be 
preserved in $\xit^R_N$. This intuition is articulated in the following lemma.

\begin{lemma}
\label{L8} We have
\boe
\label{eq32}
\lim\limits_{n\to\infty} P\bigg(\eta_N(n)_0=1\bigg)
\le\lim\limits_{t\to\infty} P\bigg(\xit^R_{N}(t)_0=1\bigg).
\eoe
\end{lemma}
\begin{proof}{}
We prove~(\ref{eq32}) in two steps:
\boe
\label{eq68}
\begin{array}{ll}
\mbox{(a) }&
\lim\limits_{n\to\infty} P\bigg(\eta_N(n)_0=1\bigg)
\le \lim\limits_{j\to\infty} P\bigg(\eta_{N}(n(j))_0=1\bigg),\\
\mbox{(b) }&
\lim\limits_{j\to\infty} P\bigg(\eta_{N}(n(j))_0=1\bigg)
= \lim\limits_{t\to\infty} P\bigg(\xit^R_{N}(t)_0=1\bigg).
\end{array}
\eoe

We prove (b) first. We write
$$
P(\xit^{R}_{N}(t)_0=1)=\sumu\limits_{j=0}^{\infty}
P\bigg(\eta_{N}(n(j))_0=1,\,\mbox{and }\taut_j\le t<\taut_{j+1}\bigg).$$
The random variable $\eta_{N}(n(j))_0$ is independent
of $\taut_j$ and $\taut_{j+1}$. Hence
$$
\bigg|P(\xit^{R}_{N}(t)_0=1)
-\lim\limits_{j\to\infty}P\bigg(\eta_{N}(n(j))_0=1\bigg)\bigg|$$
$$
=
\bigg|
\sumu\limits_{j=0}^{\infty}
\left\{P\bigg(\eta_{N}(n(j))_0=1\bigg)P(\taut_j\le t<\taut_{j+1})\right\}
-\lim\limits_{j\to\infty}P\bigg(\eta_{N}(n(j))_0=1\bigg)
\bigg|$$
$$
\le
\sumu\limits_{j=0}^{\infty}
\Bigg|P\bigg(\eta_{N}(n(j))_0=1\bigg)
-\lim\limits_{j\to\infty}P\bigg(\eta_{N}(n(j))_0=1\bigg)\bigg|\times\qquad\qquad\qquad$$
$$\qquad\qquad\qquad\qquad\qquad\times P(\taut_j\le t<\taut_{j+1})
\to0,\qquad\mbox{as }t\to\infty,$$
becauce $\taut_j\to\infty$, as $j\to\infty$ in probability.
This proves (b). 

For (a), we write 
$$
p_k=\lim_{n \to \infty} P\left(\sum_{i \in \Lambda(N)} \eta_N(n)_i =k\right),
$$
and 
$$
q_k=\lim_{j \to \infty} P\left(\sum_{i \in \Lambda(N)} \eta_N(n(j))_i =k\right).
$$
It is then clear that
$$
P(\eta_N(n)_0=1)=\sum_{k=0}^N \frac{kp_k}{N}
$$
and 
$$
P(\eta_N(n(j))_0=1)=\sum_{k=0}^N \frac{kq_k}{N}.
$$
Now observe that when there are $k<N$ vertices with fitness 1, the number of trials before 
we select a 
vertex with fitness 0 has a geometric distribution with parameter $(N-k)/k$, hence the 
expected number 
of trials is equal to
$N/(N-k)$. It follows that for $k <N$,
$$
q_k =\frac{\frac{N}{N-k} p_k}{\sum_{l=0}^{N-1} \frac{N}{N-l} p_l + (N+1)p_N}
$$
and similarly
$$
q_N =\frac{(N+1)p_N}{\sum_{l=0}^{N-1} \frac{N}{N-l} p_l + (N+1)p_N}.
$$
We write $a_k= k/N$, $b_k=N/(N-k)$ for $k < N$, and $b_N=N+1$. Then
\begin{eqnarray*}
P(\eta_N(n(j))_0=1) &=& \sum_{k=0}^N \frac{kq_k}{N} \\
&=& \frac{\sum_{k=0}^{N-1} \frac{k}{N} \frac{N}{N-k} p_k + (N+1)p_N}{\sum_{l=0}^N b_l 
p_l}\\
&=& \frac{\sum_{k=0}^N a_kb_k p_k}{\sum_{l=0}^N b_lp_l}.
\end{eqnarray*}
Now observe that for any probability vector $p_0, \ldots, p_N$, and non-decreasing 
sequences
$0 \leq a_0 \leq \cdots \leq a_N$ and $0 \leq b_0 \leq \cdots \leq b_N$, we have
$$
\left(\sum_{k=0}^N a_k p_k\right)\left(\sum_{k=0}^N b_k p_k\right)\leq \sum_{k=0}^N a_kb_k 
p_k,
$$
which can be proved by induction.
Applying this general fact to the $a_k$'s, $b_k$'s and $p_k$'s above, we find that the 
last quotient is 
bounded below by $\sum_{k=0}^N a_k p_k$ which is just $P(\eta_N(n)_0=1)$. (Note that here 
we have used the fact that we skip $N$ choices if all vertices have fitness 1: this makes 
the sequence $(b_k)$ increasing.) This proves (a).
\end{proof}

The next step in the proof of Theorem~\ref{P1}
is the following lemma, where we relate the process $\xit^R_N(t)$
to the graphical representation $\GRt$. To simplify notations 
further we will write $\GRt_N$ instead of $\GRt|_{[-N',N'']}$.

\begin{lemma}
\label{L9}
For any $t_1,t_2\ge0$, $x,y\in[-N',N'']$, if $\xit^R_{N}(t_1)_x=0$ and
$(x,t_1)\tot(y,t_2)$ in~$\GRt_N$, then $\xit^R_{N}(t_2)_y=0$.
\end{lemma}

\begin{proof}{}
Let $(x,t_1)\tot(y,t_2)$ in~$\GRt_N$,
with trace $(x_0,s_0),\dots,(x_n,s_n)$, i.e. every pair
$(x_i,s_i)$, $(x_{i+1},s_{i+1})$ is connected either by
a time segment or by an arrow. The statement 
will follow by induction, if we prove that
$x_i\in\xit^R_N(s_i)$ implies $x_{i+1}\in\xit^R_N(s_{i+1})$, 
for every $0\le i\le n-1$.
Let $x_i\in\xit^R_N(s_i)$. If $(x_i,s_i)$ 
is connected to $(x_{i+1},s_{i+1})$ by a time segment, then 
there are no symbols `$\ast$' on $(x_i,t)$, 
$t\in[s_i,s_{i+1})$. Hence, there are no arrivals at 
the time interval $[s_i,s_{i+1})$
at $\Pi_{x_i}^{\sigma_1,1,\sigma_3}$,
$\Pi_{x_i-1}^{\sigma_1,\sigma_2,1}$ and
$\Pi_{x_i+1}^{1,\sigma_2,\sigma_3}$,
$(\sigma_1,\sigma_2,\sigma_3)\in\{0,1\}^3$. 
Thus, $x_{i+1}=x_i\in\xit^R_N(s_{i+1})$,
because only the above arrivals can delete $x_i$
from $\xit^R_N(s_{i+1})$.
If $x_{i+1}=x_i+1$ and $(x_i,s_i)$ 
is connected to $(x_{i+1},s_{i+1})$ by an arrow,
then there is an arrival at
$\Pi^{\sigma_1,\sigma_2,0}_{x_i}$, for some $\sigma_1,\sigma_2\in\{0,1\}$ 
at time $s_i=s_{i+1}$, and hence, $x_{i+1}\in\xit^R_N(s_{i+1})$.
Similary, if $x_{i+1}=x_i-1$ and $(x_i,s_i)$ 
is connected to $(x_{i+1},s_{i+1})$ by an arrow,
then there is an arrival at
$\Pi^{0,\sigma_2,\sigma_3}_{x_i}$, for some $\sigma_2,\sigma_3\in\{0,1\}$ 
at time $s_i=s_{i+1}$, and again, $x_{i+1}\in\xit^R_N(s_{i+1})$.
\end{proof}

The last two lemmas imply that we have reduced the problem to $GR$. Therefore, in the next 
section, 
we 
work in this graphical representation.

\section{Proof of Theorem~\ref{P1}}

To simplify our notation $\xit^{(A,t)}_B(s)$, 
we will skip the upper index $(A,t)$, if $A=B$ and $t=0$.
We will also skip the lower index $B$, if $B=\ZZ$. 
So for example, we will write $\xit^{(A,t)}(s)$
instead of $\xit^{(A,t)}_{\ZZ}(s)$,  $\xit_{[x,\infty)}(t)$
instead of $\xit^{([x,\infty),0)}_{[x,\infty)}(t)$, and
$\xit_{(-\infty,x]}(t)$ instead of 
$\xit^{((-\infty,x],0)}_{(-\infty,x]}(t)$.
The idea to couple by a graphical representation 
the processes with various lower indices is
taken from~\cite{Sh}.

Note that for any $t\ge 0$, with probability one, 
$\xit_{[-N',\infty)}(t)\ne\emptyset$ and 
$\xit_{(-\infty,N'']}(t)\ne\emptyset$. Thus we can 
define $\llt_{-N'}(t)$ and $\rrt_{N''}(t)$ as 
the leftmost and rightmost 0's
of the processes $\xit_{[-N',\infty)}(t)$ 
and $\xit_{(-\infty,N'']}(t)$, respectively. 
The following lemma was inspired by inequality (5.2) 
in~\cite{Sh}.

\begin{lemma}
\label{L17}
$$
\mbox{(a) }\xit_{[-N',\infty)}(t)\supseteq [\llt_{-N'}(t),\infty)
\cap\xit(t),\,t\ge0,$$
$$
\mbox{(b) }\xit_{(-\infty,N'']}(t)\supseteq (-\infty,\rrt_{N''}(t)]
\cap\xit(t),\,t\ge0.$$
\end{lemma}
\begin{proof}{}
Let $y\in[\llt_{-N'}(t),\infty)\cap\xit(t)$.
Then there exists $x_1\in\ZZ$ and an open path 
$\gamma_1(s)|_0^t$ in $\GRt$ such that 
$(x_1,0)$ is connected to $(y,t)$ by $\gamma_1(s)|_0^t$.
Since $\llt_{-N'}(t)\in\xit_{[-N',\infty)}(t)$
there exists $x_2\in\ZZ\cap[-N',\infty)$
and an open path $\gamma_2(s)|_0^t$, 
laying completely within $[-N',\infty)\times\RR$
such that $(x_2,0)$ is connected to 
$(\llt_{-N'}(t),t)$ by $\gamma_2(s)|_0^t$.
Let $s^*$ be defined as
$$
s^*=\inf\{s\ge0\,:\,\gamma_2(s)\le\gamma_1(s)\}.$$
Note that by definition any open path is a {\it cadlag}
function of time, thus $\gamma_2(s^*)\le\gamma_1(s^*)$.
If $s^*=0$ then the open path $\gamma_1(s)|_0^t$ lays
completely within $[-N',\infty)$.
If $s^*>0$, we define the open path 
$\gamma_3(s)|_0^t$ as
$$
\gamma_3(s)=
\left\{
\begin{array}{ll}
\gamma_2(s),\qquad s\in[0,s^*),\\
\gamma_1(s),\qquad s\in[s^*,t],
\end{array}\right.
$$
see Figure~\ref{f6}.
The open path $\gamma_3(s)|_0^t$ has 
endpoint $(y,t)$ and lays completely 
within $[-N',\infty)\times\RR$.
Thus $(y,t)\in\xit_{[-N',\infty)}(t)$.

\fig{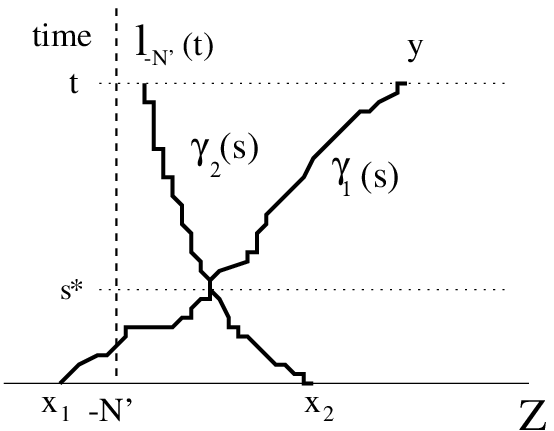}{The open path connecting
$(x_2,0)$ to $(\gamma_2(s^*),s^*)$ to $(y,t)$ 
lays completely within $[-N',\infty)\times\RR$.}{f6}

The statement in (b) can be proved in a similar way.
\end{proof}

Define $[\llt,\rrt]_N(t)$ as
$$
[\llt,\rrt]_N(t)=\xit_{[-N',\infty)}(t)\cap\xit_{(-\infty,N'']}(t), \qquad t\ge0.$$
\begin{lemma}
\label{L18}
If $[\llt,\rrt]_N(t)\ne\emptyset$, for all $t\in[t_1,t_2]$,
then for any $y\in[\llt,\rrt]_N(t_2)$ there exists
an open path $\gamma(t)|^{t_2}_{t_1}$ with $\gamma(t_2)=y$ 
and $\gamma(t)\in[\llt,\rrt]_N(t)$, $t\in[t_1,t_2]$.
\end{lemma}
\begin{proof}{}
Let $A'$ be the subset of $t^* \in [t_1,t_2]$ which have that for all
 $y\in[\llt,\rrt]_N(t^*)$,
there exists an open path $\gamma(s)|^{t^*}_{t_1}$
with $\gamma(t^*)=y$ 
and $\gamma(t)\in[\llt,\rrt]_N(t)$, $t\in[t_1,t^*]$. 
If $[\llt,\rrt]_N(t)\ne\emptyset$, for $t\in[t_1,t_2]$
then $\llt_{-N'}(t),\rrt_N(t)\in[-N',N'']$, for all $t\in[t_1,t_2]$.
Thus the event $t^*\in A'$ is determined only by the~$\GRt$
inside $[-N',N'']\times[t_1,t_2]$. 

Suppose $t^*\in A'$ and $\taut$
is the first arrival of $\Pi(N)$ after time $t^*$.
Then due to the definition of an open path 
$[t^*,\taut)\in A'$. We will prove that $\taut\in A'$.
Then, by induction, $t_2\in A'$, because there are 
only finitely many arrivals in $\Pi(N)$ at time 
interval $[t_1,t_2]$ and $t_1\in A'$. 
Let $y\in[\llt,\rrt]_N(\tau)$.
Then there exist open paths 
$\gamma_1(t)|^{\taut}_{t_1}\in[-N',\infty)$ and
$\gamma_2(t)|^{\taut}_{t_1}\in(-\infty,N'']$ with
endpoint $(y,\taut)$. Since $\llt_{-N'}(t^*)$ is the 
leftmost point of $\xit_{[-N',\infty)}(t^*)$,
we have $\gamma_1(t^*)\ge\llt_{-N'}(t^*)$.
Similarly, $\gamma_2(t^*)\le\rrt_{N''}(t^*)$.
The distance $|\gamma_1(t^*)-\gamma_2(t^*)|\le1$,
because there is only one arrival of $\Pi(N)$ in the time
interval $(t^*,\taut]$. It follows from the above 
that $\gamma_1(t^*)\in[\llt,\rrt]_N(t^*)$ or
$\gamma_2(t^*)\in[\llt,\rrt]_N(t^*)$. Suppose
$\gamma_1(t^*)\in[\llt,\rrt]_N(t^*)$. Then 
$\gamma_1(t)\in[\llt,\rrt]_N(t)$, $t\in[t^*,\taut]$.
Since $t^*\in A'$ there exists an open path 
$\gamma(s)|^{t^*}_{t_1}$ with $\gamma(t^*)=\gamma_1(t^*)$ 
and $\gamma(t)\in[\llt,\rrt]_N(t)$, $t\in[t_1,t^*]$.
Hence, the path $\gamma_3(t)|^{\taut}_{t_1}$,
defined as
$$
\gamma_3(t)=
\left\{
\begin{array}{ll}
\gamma(t),\qquad t\in[t_1,t^*),\\
\gamma_1(t),\qquad t\in[t^*,\taut].
\end{array}\right.
$$
has endpoint $(y,\taut)$, and satisfies
$\gamma_3(t)\in[\llt,\rrt]_N(t)$, $t\in[t_1,\tau]$.
See Figure~\ref{f7} for an illustration. Thus, $\taut\in A'$. The case 
$\gamma_2(t^*)\in[\llt,\rrt]_N(t^*)$ can be 
done similary.

\fig{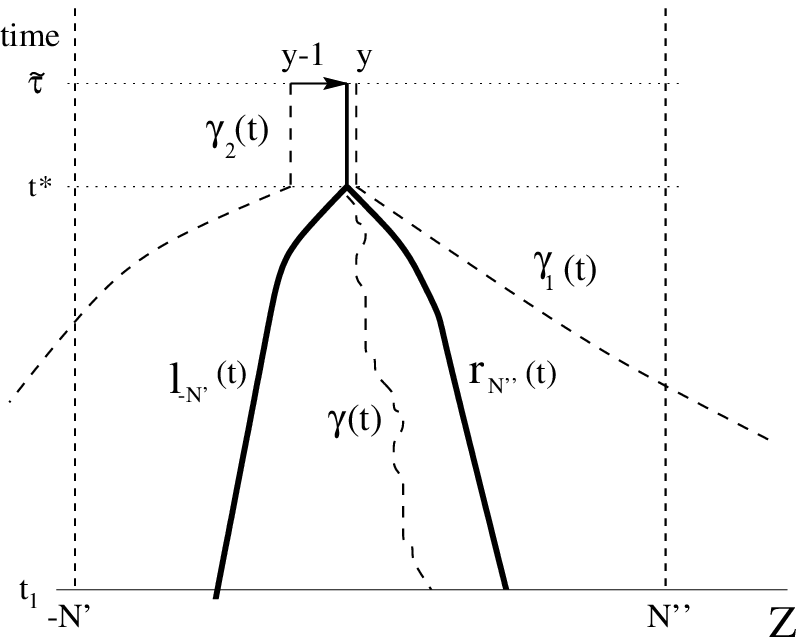}{At the realisation 
$[\llt,\rrt]_N(\taut)=[\llt,\rrt]_N(t^*)=\{y\}$.
Since $(y,t^*)$ is connected to $(y,\taut)$ by
$\gamma(t)|^{\taut}_{t^*}$ and $t^*\in A'$,
there exists an open path with  endpoint 
$(y,\taut)$, laying completely within 
$[\llt,\rrt]_N(t)$, $t\in[t_1,\taut]$.}{f7}

\end{proof}

We need a number of results which are very similar to the corresponding results for the 
standard contact 
process. Therefore, we omit the proofs of the following three lemmas. Their proofs are modifications 
of the proofs of Theorem 3.19, Theorem 3.21 and Corollary 3.22 in \cite{L}. Note 
that in these three 
lemmas, we require $q$ to be close enough to one. It is at this point where the rescaling 
of time by 
a factor $1/(1-q^3)$ comes in. With this rescaling of time, the intensity of arrows gets 
large when 
$q$ gets close to one.  The (omitted) proofs of the three lemmas involve comparison with 
oriented 
percolation, and in order to make sure that the appropriate oriented percolation model 
percolates, we need a high intensity of arrows. 

\begin{lemma}
\label{L1}
If $q$ is close enough to one, then there exists $\nu(q)>0$
such that for any $t>0$
\boe
P\bigg(\xit(t)_0=0\bigg)>\nu(q).
\eoe
\end{lemma}

\begin{lemma}
\label{L2}
Let $x\in\ZZ$ and let $\llt_x(t)$ be the leftmost zero of
$\xit_{[x,\infty)}(t)$, and $\rrt_x(t)$ be the rightmost zero of
$\xit_{(-\infty,x]}(t)$.
If $q$ is close enough to one, then there exist 
$c_1(q),c_2(q)>0$, depending only on $q$, such that 
for any $m\in\NN$ and $t>5m^2$
\boe
P\bigg(\llt_x(s)>x+m,\mbox{ for some }s\in[t-5m^2,t]\bigg)<c_1(q)e^{-c_2(q)m},\\
P\bigg(\rrt_x(s)<x-m,\mbox{ for some }s\in[t-5m^2,t]\bigg)<c_1(q)e^{-c_2(q)m}.
\eoe
\end{lemma}

\begin{lemma}
\label{L3}
For $q$ close enough to one there exists
$c(q)>0$, depending only on $q$ such that
$$
P\left(
\exists t'\in[0,N/2]\,:\,(0,0)\tot(N,t')\mbox{ in }\GRt_{[0,\infty)}
\right)>c(q).
$$
\end{lemma}

\medskip\noindent
Now we partition the time axis into intervals of length $N$, and we call the interval 
$[iN,(i+1)N)$ 
the $i$-th {\em level}.

\begin{definition} 
We call level $i$ {\rm normalising} if there exist
$t,t',t''\in[iN,(i+1)N)$ and $x\in[-N',N'']$ such that
$\xit^R_{N}(t)_x=0$, $(x,t)\tot(x-N,t')$ in $\GRt_{(-\infty,x]}$ 
and $(x,t)\tot(x+N,t'')$ in $\GRt_{[x,\infty)}$.
\end{definition}
See Figure~\ref{f3} for an illustration of this definition.

\fig{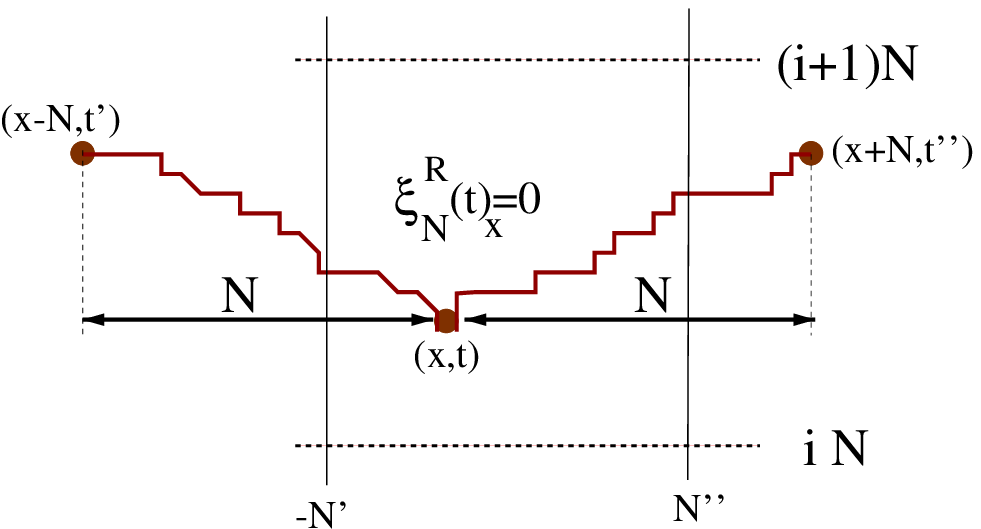}{The i-th level is normalising.}{f3}

We will use normalising levels to connect different open paths in $GR$. For this to work, 
we have to 
make sure that there are enough normalising levels. This is the content of the following 
key lemma.

\begin{lemma}
\label{L4}
For $q$ close enough to one, 
$N$ large enough, and $T>N^2+N$, there exists
$c(q)>0$, depending only on $q$, such that
the probability to find no normalising level among levels
$\lfloor T/N\rfloor-N,\dots,\lfloor T/N\rfloor$
is at most $e^{-c(q)N}$.
\end{lemma}
\begin{proof}{}
The events
$$
\bigg\{i\mbox{-th level is normalising}\bigg\},
\qquad i\in\NN
$$
are not independent. But we will construct independent events which guarantee that certain 
levels 
are normalising. To do this carefully, let ${\cal F}_i(N)$ be the $\sigma$-algebra 
generated by the 
restriction of $\GRt$ to $\ZZ\times[0,iN+N)$. Let 
$$
t^*_i=\min\bigg\{iN+N/2,
\inf\{t\ge iN\,:
\,\xit^R_N(t)\cap[-N',N'']\ne\emptyset\}\bigg\},
$$
and
$$
x^*_i=\left\{
\begin{array}{rl}
\min\{x\in[-N',N'']\,:
\,\xit^R_{N}(t_i^*)_x=0\},
&\mbox{ if }t^*_i<iN+N/2,\\
0,&\mbox{ if }t^*_i=iN+N/2.
\end{array}\right.$$
Let $A_i(N)$ be the event
that $(x^*_i,t^*_i)\tot(x^*_i+N,t')\mbox{ in }\GRt_{[x^*_i,\infty)}$, 
and $(x^*_i,t^*_i)\tot$
$(x^*_i-N,t'')\mbox{ in }\GRt_{(-\infty,x^*_i]}$, 
for some $t',t''\in[t^*_i,t^*_i+N/2)$. Note that $A_i(N)\cap\{t^*_i<iN+N/2\}$ implies that 
the 
$i$-th level is normalising. Since $t^*_i$ is a stopping time and because of 
Lemma~\ref{L3},
there exists $p_{q,A}>0$, depending only on $q$, such that
\boe
\label{eq9}
P\bigg(A_i(N)\bigg)\ge p_{q,A},
\eoe
uniformly in $N$. Observe that $A_i(N)$ is ${\cal F}_i(N)$-measurable, and is independent 
of
${\cal F}_{i-1}(N)$. 

We are now going to make sure that $\{t^*_i<iN+N/2\}$ occurs often enough.
Let $s_i^*$ be defined as the smallest element $s$ of the $i$-th level for which one of 
the 
following three conditions is satisfied:

\medskip\noindent
(1) $\xit_N^R(s)=\emptyset$;\\
(2) $\xit_N^R(s) \cap [-N',N''] \neq \emptyset$;\\
(3) $s= iN + N/4$.\\

\noindent
Note that $s_i^*$ is a stopping time (with respect to the natural filtration) and that $s_i^* \leq iN + 
N/4$. 
We will now give a condition in terms of $\GRt$ within the $i$-th level which ensures that 
$s^*_i<iN 
+ N/4$. 

If $\xit_N^R(iN)=\emptyset$ or $\xit_N^R(iN) \cap [-N',N''] \neq \emptyset$, then 
$s_i^*=iN$. The 
remaining cases are those where $\xit_N^R(i)$ is not empty and a subset of $\{-N'-1, 
N''+1\}$. We 
now {\em pretend} that $\xit_N^R(iN)$ is not empty and a subset of $\{-N'-1, N''+1\}$, 
giving three 
possible situations,
namely $\xit_N^R(iN)=\{-N'-1\}$, $\xit_N^R(iN)=\{N''+1\}$ or 
$\xit_N^R(iN)=\{-N'-1,N''+1\}$. In each of 
these situations, we compute 
$$
\inf\bigg\{s\ge0: \xit^R_N(iN+s)=\emptyset
\mbox{ or }\xit^R_N(iN+s)\cap[-N',N'']\ne\emptyset\bigg\},
$$
which we denote by $S^1, S^2$ and $S^3$ respectively. 
We denote the maximum of these three numbers by $S_i^*$:
$$
S_i^*=\max\{S^1, S^2, S^3\},
$$
and define the event $B_i(N)$ as 
$$
B_i(N)=\{S^*_i<N/4\}.
$$
Note that $B_i(N)$ is measurable with respect to the $\sigma$-algebra generated
by $GR$ restricted to the $i$-th level, and therefore the $B_i(N)$'s are mutually 
independent for 
different $i$'s. Also observe that occurrence of $B_i(N)$ implies that $s_i^* < N/4$, and 
that 
\begin{equation}
\label{charlotte}
\lim_{N \to \infty} P\bigg(B_i(N)\bigg)= 1.
\end{equation}
Next, define $C_i(N)$ as the event that the $(N+1)$-th arrival of $\Pi(N)$ after time 
$s_i^*$ takes 
place before time $s_i^*+N/4$. Since $\Pi(N)$ has intensity of order $N$, the number of 
arrivals of 
$\Pi(N)$ in a time interval of length $N/4$ has a Poisson distribution with mean of order 
$N^2$. 
This implies that 
\begin{equation}
\label{hansje}
\lim_{N \to \infty} P\bigg(C_i(N)\bigg)= 1.
\end{equation}
Finally, $D_i(N)$ is defined as the event that the first arrival of $\Pi(N)$ after time $s_i^*$ is at a 
vertex in 
$[-N', N'']$, or that there is no such arrival during the time interval $[s_i^*,s_i^*+N/4)$. It 
is clear 
that
\begin{equation}
\label{annemarie}
\lim_{N \to \infty} P\bigg(D_i(N)\bigg)= 1.
\end{equation}
The events $C_i(N)$ and $D_i(N)$ are ${\cal F}_i(N)$-measurable, and independent of ${\cal 
F}_{i-1}(N)$. Now we have that
$$
\bigg\{B_i(N) \cap C_i(N) \cap D_i(N)\bigg\} \subseteq \bigg\{t_i^* < iN + N/2\bigg\},
$$
and therefore
$$
\bigg\{A_i(N) \cap B_i(N) \cap C_i(N) \cap D_i(N)\bigg\} \subseteq \bigg\{i\mbox{-th level 
is 
normalising}\bigg\}.
$$
Also, for $N$ large enough, we have, according to (\ref{eq9}), (\ref{charlotte}), 
(\ref{hansje})
and (\ref{annemarie}) that 
$$
P\bigg(A_i(N) \cap B_i(N) \cap C_i(N) \cap D_i(N)\bigg) > c_1(q),
$$
for some $c_1(q)>0$, uniformly in $N$. We may now write, using the independence of all 
events
$$
P\bigg(\mbox{none of the levels } 
\lfloor T/N\rfloor-N,\dots,\lfloor T/N\rfloor
\mbox{ are normalising }\bigg)
$$
\begin{eqnarray*}
&\le& P\bigg(\bigcap_{i=\lfloor T/N\rfloor-N}^{\lfloor T/N\rfloor}
\left(A_i(N) \cap B_i(N) \cap C_i(N) \cap D_i(N)\right)^c\bigg)\\
&=& \prod_{i=\lfloor T/N\rfloor-N}^{\lfloor T/N\rfloor}
P\bigg(\left(A_i(N) \cap B_i(N) \cap C_i(N) \cap D_i(N)\right)^c \bigg) \\
&\le& e^{-c(q)N},
\end{eqnarray*}
for some $c(q) >0$.
\end{proof}

\begin{proof}{ of Theorem~\ref{P1}}
Due to the symmetry of BS-process we can work with $i=0$.
%W.l.g. we assume $i=0$.
According to Lemma~\ref{L8}, it suffices to prove that for
$q$ close enough to one, there exists $c_q>0$ 
such that for any $N$ sufficiently large,
$$
\lim\limits_{t\to\infty}P(\xit^R_{N}(t)_0=0)\ge c_q.
$$
For any $T>0$ and $N$ we have
\boe
\label{eq1}
P\bigg(\xit^R_N(T)_0=0\bigg)
\ge P\bigg(\xit(T)_0=0\bigg)-P\bigg(\xit^R_N(T)_0=1,\,\xit(T)_0=0\bigg).
\eoe
The first term in~(\ref{eq1}) is independent of $N$ and positive,
according to Lemma~\ref{L1}. 
Thus it is remains to prove that 
\boe
\label{eq2}
P\bigg(\xit^R_N(T)_0=1,\,\xit(T)_0=0\bigg)\to 0,
\mbox{as }N\to\infty,\mbox{ uniformly in }T>T(N),
\eoe
for some $T(N)<\infty$.
Let $T>N^2+N$. According to Lemma~\ref{L2} and the stationarity 
of $\GRt$, for $q$ close enough to one, 
there exists $c_1=c_1(q),c_2=c_2(q)>0$ depending only on $q$, such that 
\boe
\label{eq3}
\begin{array}{ll}
( i)&P\bigg(\llt_{-N'}(t)>0,\mbox{ for some }t\in[T-N^2-N,T]\bigg)
<c_1 e^{-c_2 N},\\
(ii)&P\bigg(\rrt_{N''}(t)<0,\mbox{ for some }t\in[T-N^2-N,T]\bigg)
<c_1 e^{-c_2 N}.\\
\end{array}
\eoe
It follows from~(\ref{eq3}) and Lemma~\ref{L4} that
$$
P\bigg(\xit^R_N(T)_0=1,\,\xit(T)_0=0\bigg)$$
\begin{equation}
\label{eq4}
\begin{array}{l}
\le P\left(
\begin{array}{c}
\xit^R_N(T)_0=1,\,\xit(T)_0=0,\\
\llt_{-N'}(t)\le0\le\rrt_{N''}(t),\mbox{ for all }t\in[T-N^2-N,T],\\
\exists i\in\{\lfloor T/N\rfloor-N,\dots,\lfloor T/N\rfloor \}
\,:\,i\mbox{-th level is normalising}
\end{array}
\right)\\
\qquad + c_1 e^{-c_2 N}
\end{array}
\end{equation}

\fig{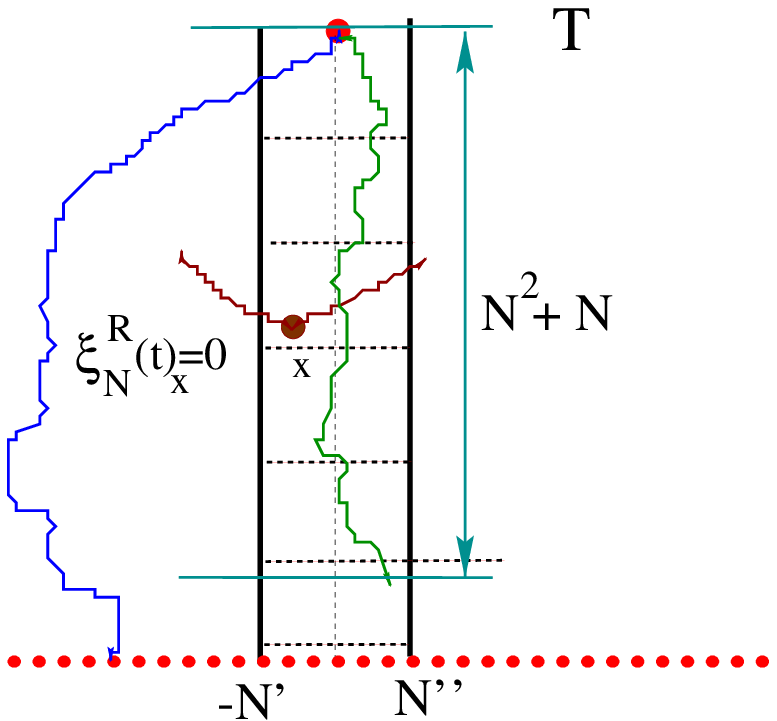}{The main idea of the proof. With probability close to one,
$\{\xit(T)_0=0\}$ implies $\{\xit^R_N(T)_0=0\}$.}{f2}

Consider the first term of~(\ref{eq4}).
It follows from Lemma~\ref{L17} that the event
$$
\bigg\{\llt_{-N'}(T)\le0\le\rrt_{N''}(T),\,\xit(T)_0=0\bigg\}$$
implies that $\{0\}\in[\llt,\rrt]_N(T)$. Then, by Lemma~\ref{L18}, 
the event
$$
\left\{
\begin{array}{c}
\{0\}\in[\llt,\rrt]_N(T),\\
\llt_{-N'}(t)\le\rrt_{N''}(t),\mbox{ for all }t\in[T-N^2-N,T],
\end{array}
\right\}$$
implies that there exists an open path $\gamma(t)|_{T-N^2-N}^{T}$ with endpoint $(0,T)$ 
and laying completely within $[-N',N'']$. If the
$i$-th level is normalising then, by definition, there exists 
$t,t',t''\in[iN,(i+1)N)$, $x\in[-N',N'']$ and open paths
$\gamma_l(s)|_t^{t'}:(x,t)\tot(x-N,t')$ in $\GRt_{(-\infty,x]}$, 
$\gamma_r(s)|_t^{t''}:(x,t)\tot(x+N,t'')$ in $\GRt_{[x,\infty)}$. 
Either $\gamma_l(s)|_t^{t'}$ or $\gamma_r(s)|_t^{t''}$
intersects $\gamma(t)|_{T-N^2-N}^{T}$, see Figure~\ref{f2}. 
Hence $(x,t)$ is connected to $(0,T)$ by an open path in $\GRt_N$,
and by Lemma~\ref{L9} we have $\xit^R_N(T)_0=0$. Hence the first 
term at the right hand side of (\ref{eq4}) equals zero and the second term gives us
the theorem.
\end{proof}

\medskip\noindent
{\bf Acknowledgement:} We would like to thank Jeff Steif 
and J\'er\'emy Barbay for helpful discussions.

\end{document}